\documentstyle[11pt,aaspp,psfig]{article}
\def\gs{\mathrel{\raise0.35ex\hbox{$\scriptstyle >$}\kern-0.6em \lower0.40ex\hbox{{$\scriptstyle \sim$}}}}
\def\ls{\mathrel{\raise0.35ex\hbox{$\scriptstyle <$}\kern-0.6em \lower0.40ex\hbox{{$\scriptstyle \sim$}}}}

\begin{document}
\small
{\hfill \fbox{{\sc ApJ in press}}}

\title{Star formation and selective dust extinction in 
luminous starburst galaxies}
\author{
Bianca M.\ Poggianti,$^{\!}$\altaffilmark{1}
Alessandro Bressan,$^{\!}$\altaffilmark{1}
Alberto Franceschini$^{\!}$\altaffilmark{2}
}
\smallskip

\affil{\scriptsize 1) Osservatorio Astronomico di Padova, vicolo dell'Osservatorio 5, 35122 Padova, Italy.}
\affil{\scriptsize 2) Dipartimento di Astronomia, vicolo dell'Osservatorio 5,
35122 Padova, Italy.}

\begin{abstract}
We investigate the star formation and dust extinction properties
of very luminous infrared galaxies whose spectra display a strong 
$\rm H\delta$ line in absorption and a moderate [O{\sc ii}]
emission (e[a] spectrum). This spectral combination has been suggested
to be a useful method to identify dusty starburst galaxies at any redshift
on the basis of optical data alone.
We compare the average e(a) optical spectrum 
with synthetic spectra that include both the stellar and the nebular
contribution, allowing dust extinction to affect differentially the
stellar populations of different ages. 
We find that reproducing the e(a) spectrum 
requires the youngest stellar generations to be significantly more
extinguished 
by dust than older stellar populations, and implies a strong ongoing
star formation activity at a level higher than in quiescent spirals.
A model fitting the optical spectrum does not necessarily produce
the observed FIR luminosity and this can be explained by the existence of
stellar populations which are practically obscured at optical
wavelengths. Models in which dust and stars are uniformly mixed
yield a reddening of the emerging emission lines which is too low
compared to observations: additional foreground reddening is required.
\end{abstract}

\keywords{galaxies:evolution -- galaxies: starburst -- dust:extinction --
infrared:galaxies}

\sluginfo

\section{Introduction}

The effects of dust extinction are crucial for 
interpreting the spectra of nearby and distant galaxies, particularly 
after that new facilities from the ground (e.g. {\it SCUBA} on 
{\it JCMT}) and from 
space ({\it ISO}) have established that 
the IR emissivity of galaxies was much enhanced in the past with regard 
to what is typically observed in local galaxy populations
(Elbaz et al. 1999; Barger et al. 2000, Franceschini et al. 2000, 
Ivison et al. 2000, Smail et al. 2000).
Further evidence about the very important role played by dust reprocessing 
during the evolution of galaxies was provided by the
detection of a diffuse far-IR background likely originating from
distant and primeval galaxies, whose
dominance over the optical background testifies
hidden star formation at unexpectedly high rates
(Puget et al. 1996; Dwek et al. 1998).

Waiting for future spectroscopic capabilities in the far-IR which will be 
offered by a variety of missions (SIRTF, FIRST, and NGST), 
only indirect clues about the relation between
dust and stellar population emissions can be gained from optical spectra.

In the local universe, a large fraction of the FIR-luminous galaxies
are characterized by a peculiar combination of spectral features in the
optical: a strong $\rm H\delta$ line in absorption (EW$> 4$ \AA) and a moderate
[O{\sc ii}] emission (EW$< 40$ \AA) (Poggianti \& Wu 2000, hereafter PW00). 
Galaxies with this type of spectra were named ``e(a)'' galaxies
\footnote{We stress that these are {\it not} the so-called
``E+A'' (or k+a) spectra which \it by definition \rm 
have \it no detectable emission lines \rm (Dressler \& Gunn 1983),
at least at the level of the [O{\sc ii}] detection limit of the 
spectroscopic surveys of distant galaxies (typically 3 \AA). Nevertheless,
it is possible that some of the k+a spectra are e(a) galaxies with
an extremely low [O{\sc ii}] (P99, Smail et al. 1999).}
and were found to be quite numerous in the cluster and
field environments at $z=0.4-0.5$ (Dressler et al. 1999, Poggianti
et al. 1999, hereafter P99). The equivalent width of their $\rm H\delta$ line
exceeds that of typical, quiescent spirals at low-z and their
low [O{\sc ii}]/$\rm H\alpha$ ratios are consistent with
the emission line fluxes being highly extincted by dust (P99, PW00).

Interestingly, the combination of moderate emission lines and
strong early Balmer lines in absorption has been subsequently found
also in the first spectroscopic surveys of the high-redshift
galaxy populations detected by {\it ISO} (Aussel 1999, Flores et al. 1999). 
The association between the e(a) signature and the FIR-luminous galaxies
seems to be confirmed at any redshift and
suggests that the e(a) spectra -- despite their moderate emission lines --
actually belong to highly obscured, {\it starburst} galaxies.
One of the most important pieces of information that is still missing
is the physical origin of this type of spectra: what 
star formation and dust properties produce this peculiar spectral
combination?
P99 and PW00 proposed that the e(a) spectrum originates 
from ``selective extinction'' in a galaxy
where the youngest stars in the HII regions 
are much more dust-embedded than the older stellar populations,
which are more widely distributed throughout the galaxy.
Such an age-dependent obscuration would cause the emission lines
to be highly extincted
(hence producing the low blue-to-red emission line ratios observed) and
would allow the older, A-type stars to dominate the integrated spectrum
at $\lambda \sim 4000$ \AA $\,$ (thus explaining the strong 
$\rm H\delta$ line in absorption).

The existence of a selective extinction in galaxies
is well known, but it is usually ignored when dealing with the global 
effects of dust on the spectral energy distribution of galaxies.
Stars are expected to spend the very beginning of their evolution deeply
embedded in dusty environments, later drifting away from or dispersing
the molecular clouds where they were born (Calzetti et al. 1994 and
references therein, Granato et al. 2000). As discussed in PW00, 
an age-dependent dust obscuration could explain
the numerous observational studies which measure discrepant extinction values
when using different spectral ranges/features
(Israel \& Kennicutt 1980, Viallefond et al. 1982, Caplan \& Deharveng 1986,
Fanelli et al. 1988, Bohlin et al. 1990, Keel 1993,
Calzetti et al. 1994, Veilleux et al. 1995, Lancon et al. 1996,
Mayya \& Prabhu 1996, Calzetti 1997,1998, Mas-Hesse \& Kunth 1999).
Moreover, Zaritsky (1999)
found a clear trend of decreasing mean extinction with declining stellar
effective temperature (i.e. stellar age) in the Large Magellanic Cloud.

So far, no attempt has been made to model the whole e(a) spectrum
in a quantitative way.
The aim of this work is to test whether the average e(a) spectrum
can be modelled as a current starburst assuming a selective
dust extinction that increases 
as the age of the stellar population decreases.
In the following we employ a spectrophotometric model that includes both
the stellar component and the thermal emission of the gas ionized
by young stars. Such a model will be considered successful
only if it is able to reproduce simultaneously the shape of the 
optical observed
continuum (3600-6900 \AA) and the strength of the
most important emission/absorption lines:
[O{\sc ii}]$\lambda$3727,
$\rm H\delta$ at $\lambda = 4101$ \AA, $\rm H\beta$ at $4862$ \AA $\,$ 
and $\rm H\alpha$ at $6563$ \AA.

The plan of the paper is as follows: in \S2 we present the 
spectroscopic catalog and the average observed e(a) spectrum
and in \S3 we describe the ingredients and the structure of our model.
The results are discussed in \S4, where we examine starburst 
models with selective extinction (\S4.1), post-starburst
models (\S4.2) and models with constant extinction (\S4.3).
Our main conclusions are summarized in \S5.

\section{Observations}
The observed e(a) spectra are taken from
the complete spectroscopic catalog of Very Luminous Infrared Galaxies 
(log($L_{IR}/L_{\odot}) \ge 11.5$, VLIRGs) presented in Wu et al. (1998a,b).
Among these, we selected those 19 galaxies with a secure e(a) spectrum and
a spectral coverage extending below the [O{\sc ii}] line
and beyond the $\rm H\alpha$ line. 
These are nuclear spectra covering at least the central 2 kpc of 
the galaxy (see also \S4). None of these galaxies is 
an AGN (either Seyfert1 or Seyfert2) according to the emission line ratios
(PW00).
The average e(a) spectrum is shown in Fig.~1
and the equivalent widths of the main lines are listed in Table~3.
\footnote{Spectra have not been corrected for
foreground Galactic extinction, which is
negligible (mean Galactic E(B-V) = 0.04) in comparison with the 
intrinsic extinction of these galaxies (Wu et al. 1998b).}

As mentioned in \S1, moderate emission lines and
strong early Balmer lines in absorption have been found
also in the spectra of the high-redshift IR populations. In fact, the 
average spectrum 
of the distant starburst galaxies detected at 15 $\micron$ by {\it ISO} 
in one of the fields of the Canada-France Redshift Survey (Flores et al. 
1999) turns out to be very similar to the average e(a) spectrum. 
The comparison is presented in Fig.~1 and shows that -- within the
wavelength range in common ($\sim$3700-5100 \AA) -- 
both the continuum shape
and the strength of the [O{\sc ii}], $\rm H\delta$ and $\rm H\beta$
lines are alike. 

\hbox{~}
\vspace{-1.6in}
\centerline{\psfig{file=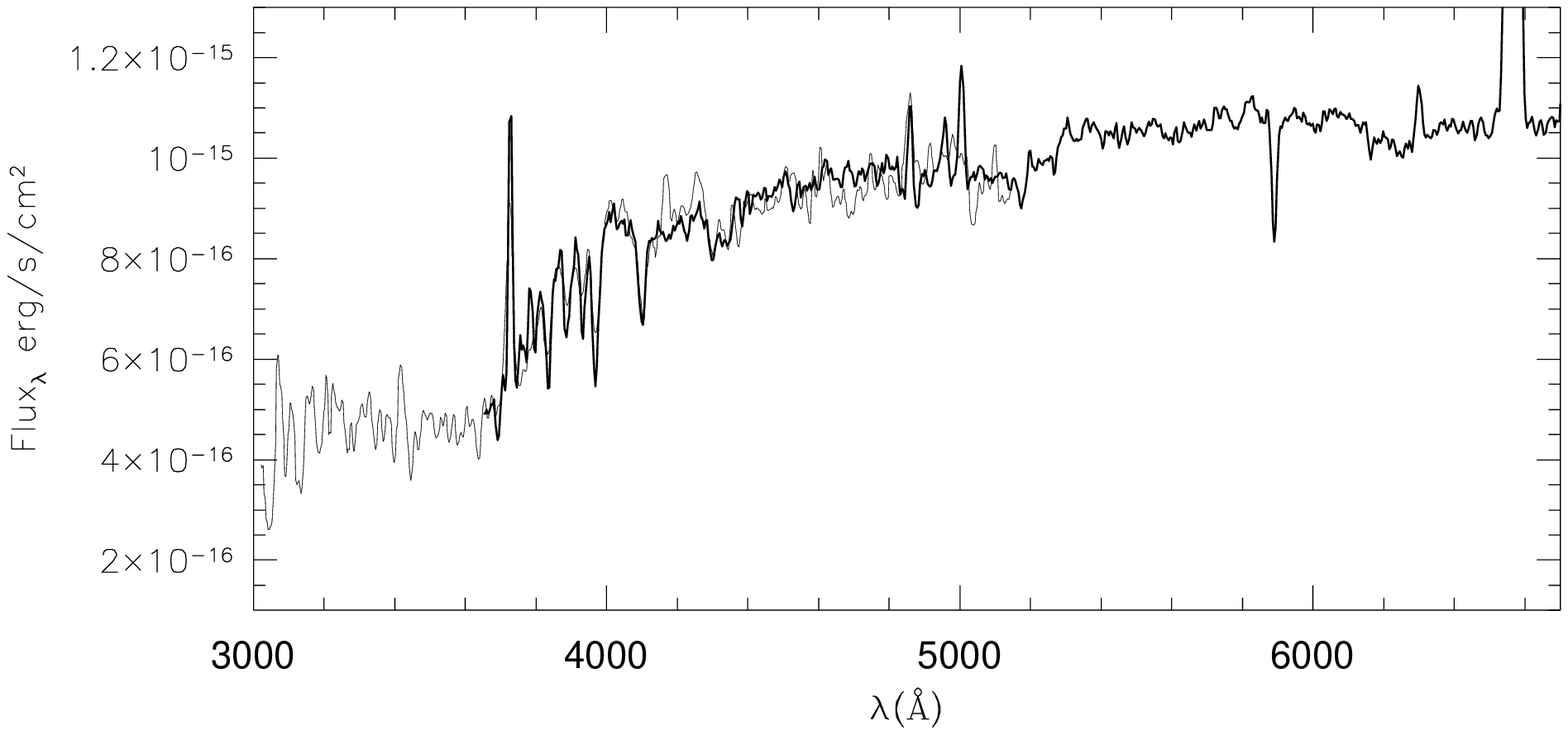,angle=0,width=4in}}
\vspace{-0.8cm}
\noindent{\scriptsize
\addtolength{\baselineskip}{-3pt} 
\hspace*{0.3cm} {\bf Fig.~1.} A rest-frame comparison of the average 
e(a) spectrum of local Very Luminous Infrared galaxies  (thick line) and 
the spectrum of the ISO starburst 
galaxies at $<z>\sim 0.5$ (thin line). 
The latter is taken from Fig.~12 in Flores et al. (1999) and 
is the average spectrum of 5 ISO-detected galaxies at $z<0.7$ whose 
spectral energy distributions at visible, near-IR, MIR and radio wavelengths
resemble those 
of highly reddened starbursts in the local universe
(Schmitt et al. 1997); it has been 
normalized to the VLIRGs spectrum over the wavelength range in common.
The spectral resolution is 10 \AA $\,$
(Wu) and 40 \AA $\,$ (Flores). 
\addtolength{\baselineskip}{3pt}
}

\section{Models}

Instead of employing a highly complex tool (i.e. a complete
spectrophotometric model), we preferred to develop and use a 
``simplified model''. This model retains all the basic physical
ingredients required by the complexity of an integrated spectrum, but
includes only
those stellar populations that are a priori essential because they 
are known to affect
the spectral features that we wish to reproduce. 

The integrated model spectrum has been generated as a combination 
of 10 stellar populations of different ages, listed in Table~1. 
Each stellar generation is born with a Salpeter IMF 
between 0.1 and 100 $M_{\odot}$.
The ages of the 10 populations have been 
chosen considering the evolutionary timescales associated with the 
observational
constraints: the youngest generations ($10^6, 3 \cdot 10^6, 8 \cdot 10^6,
10^7$ yr) are responsible for the ionizing photons that produce the 
emission lines;
the intermediate populations ($5 \cdot 10^7, 10^8, 3 \cdot10^8, 5 
\cdot 10^8, 10^9$ yr) are those with the strongest Balmer lines in 
absorption\footnote{The EW($\rm H\delta$) is always $>4$ \AA $\,$ in 
the spectra of these intermediate
populations.}, while older generations of stars (named
``old'' in Table~1 ) have been modelled as a constant star formation rate 
(SFR) between 1 and 12 Gyr before the moment of the observation
and can give a significant contribution to the spectral continuum, hence
affecting also the equivalent widths of the lines.
This approach allows us to a) reduce 
the number of parameters to a minimum: a complete spectrophotometric model 
finds the integrated spectrum summing over typically
50 single stellar populations of various ages, instead of 10;
b) explore also bursting, discontinuous star formation histories
which more likely resemble the irregular histories of ``real''
starburst galaxies.
Adopting as star formation history 
an analytic function + a single or multiple burst
would force us to restrict the parameter space investigated according to 
some a priori assumptions;
c) easily implement different extinctions for stellar populations
of different ages, interpreting the results in a 
straightforward manner.

The composite (stars+gas) spectrum of each single generation 
has been produced with the spectrophotometric code of Barbaro \&
Poggianti (1997, BP97), that includes both the stellar component and the 
thermal emission of the ionized gas. This model is fully described in BP97;
here we only summarize its salient characteristics.
Between 3500 and 7500 \AA, the stellar model
is based on the stellar spectral library of Jacoby et al. (1984) which has
enough resolution to study absorption features, such as the stellar lines
of the Balmer series in absorption. The ionizing flux of the young stellar
populations is computed using stellar atmosphere models (Kurucz 1993);
from this, the luminosities of the nebular hydrogen lines (Balmer series) are
derived assuming case B recombination (Osterbrock 1989) while the strength
of other lines -- such as [O{\sc ii}]$\lambda$3727 -- is calculated using 
HII region models (Stasinska 1990). When non-negligible (for ages $< \rm a 
\, few \, 10^7$ yr), the nebular contribution (lines + nebular continuum) 
is added to the stellar component to give the total spectrum.
For this project, both the stellar and the nebular components were found
assuming solar metallicity.
The model spectra were broadened with a Gaussian degrading the original model 
resolution ($\sim 4$ \AA) in such a way to match the resolution of the 
observed spectra ($\sim 10$ \AA).

Each single generation (SG) is assumed to be extincted by dust
in a uniform screen according to the standard extinction law of the 
diffuse medium in our Galaxy ($R_V=A_V/E(B-V)=3.1$, Cardelli et al. 1989) 
unless otherwise stated (see \S4). The extinction value E(B-V)
is allowed to vary from one stellar population to another and 
the extincted spectral energy distributions of all the single generations
are added up to give the total integrated spectrum.

Within a chosen star formation scenario (see \S4), 
the best-fit model was obtained by
minimizing the differences between selected features in the observed
and model spectra. A merit function was constructed considering 
$N=12$ features: the equivalent width of four lines
([OII]$\lambda$3727,H$\delta$, H$\beta$ and H$\alpha$) and the
relative intensities of the continuum flux in eight almost featureless
spectral windows (3770-3900\AA, 4020-4070\AA, 4150-4250\AA,
4600-4800\AA, 5060-5150\AA, 5400-5850\AA, 5950-6250\AA $\,$ and
6370-6460\AA):

\begin{equation}
{(MF)}^2 = \sum_{i=1}^N{W_i^2 (\frac{O_i-M_i}{E_i+E_0})^2}
\end{equation}

where the quantities M$_i$, O$_i$, E$_i$, E$_0$ and W$_i$
refer to the value predicted by the model, the observed value,
the accuracy, the minimum error and the weight assigned to the
$i$ feature, respectively.
An accuracy of 1\% ad 10 \% was adopted for the flux in the continuum bands
and for the equivalent widths, respectively. 
These are comparable to the observational uncertainties
if we consider that the error on the relative flux calibration
amounts to a few percent, the typical error in the measured equivalent widths
is between
10\% and 20\% and we are dealing with an average spectrum.\footnote{In 
addition, we checked that assuming a lower accuracy (4\% and 20\%
in the continuum and in the equivalent widths, respectively) does not
modify our main conclusions.}

All quantities were given a unit weight but for 
the 3770-3900 \AA $\,$ continuum and for 
the [OII]$\lambda$3727 line, to which a lower weight 
was assigned (0.4 and 0.5, respectively). 
This choice is due to the enhanced
flux calibration uncertainty of Jacoby's spectra
below 4000 \AA $\,$ and
to the fact that, unlike the Balmer
emission lines, the
intensity of the [OII]$\lambda$3727 line is not related in a simple way
to the star formation rate. In fact, the [O{\sc ii}] line
is sensitive to several other factors such as the
hardness of the ionizing spectrum and the local conditions in the
nebula, the density and the metallicity of the gas. 
We find that fitting the observed [O{\sc ii}] strength requires a 
line intensity higher than the one predicted by models
of solar metallicity 
\footnote{We note that a metallicity below solar produces a stronger 
[O{\sc ii}] line than the solar case, but causes also an
enhancement of the [OIII]$\lambda$5007 line which instead is quite
weak in the e(a) spectrum.},
unless the presence of hot Wolf Rayet stars is included. 
At ages between 3 Myr and 5 Myr these stars may enhance the
ratio [OII]$\lambda$3727/H$\alpha$ by a factor of about three (Garcia
Vargas et al. 1995), hence we chose to take this factor into account 
increasing the [O{\sc ii}] flux of the 3 Myr old single generation.

The merit function was minimized with the method of
the simulated annealing (routine AMEBSA in Numerical Recipes), which
is particularly suited to avoid the rapid convergence to a local minimum.
The free parameters (the relative intensity of the SGs and the 
extinction values)
were let to vary within specified ranges defined by the kind of
evolutionary scenario considered.

In principle, the observed FIR/optical luminosity
ratio could be treated as a further constraint of the model,
\footnote{The FIR luminosities given 
by W98a are computed from
the IRAS 60 $\micron$ and 100 $\micron$ fluxes as
$F_{IR}=1.75[2.55S_{60}+1.01S_{100}] 10^{-14} \, W \, m^{-2}$
and are approximately total FIR luminosities in the range 
1--1000 $\micron$.} 
hence, in order to predict the FIR emission, we computed a new set of
single stellar populations 
by extending the Jacoby stellar spectra below 3510 \AA $\,$ and above
7400 \AA $\,$ using Kurucz (1993) stellar atmosphere models, 
with a procedure similar to that adopted by Longhetti et al. (1999). 
Once the parameters of a best-fit model were determined, 
the FIR luminosity was evaluated as the difference between the 
total unextinguished 
spectrum and the final (extinguished) spectrum.

\section{Results}
In this section we show the results of some selected models 
whose best-fit parameters are listed in
Tables~1 and ~2. For each model, the left column (SF) gives the 
star formation rate (mass per year, after 
having normalized to SFR=1 $yr^{-1}$ in the old population) 
and the right column gives the color excess E(B-V).
The visual comparison of these models
with the average e(a) spectrum is shown in Figs.~2, 3 and 4
while Table~3 presents the equivalent widths of the four lines considered,
the FIR to optical luminosity ratio and the current star formation
rate of each model.

Different evolutionary scenarios have been examined
in order to address the following questions:

a) Can the starburst, selective extinction scenario explain the
e(a) spectra?

b) What constraints can be placed on the burst duration and strength?

c) Can the post-starburst scenario explain the e(a) spectra?

d) Can the e(a) spectrum be fitted by models with a constant value of
extinction?

\noindent
\subsection{Can the starburst selective-extinction scenario explain the 
e(a) spectra?}
At first we consider models in which the star formation rate
is roughly constant (within a factor $\sim 1.5$)
during the current starburst and much lower before that
(models A and B in Table~1 and Fig.~2).
The extinction varies with the age of the population,
being significantly higher at younger ages.

Model A is an example of a starburst that began $2 \cdot 10^8$
yr ago. The quality of the fit can be
assessed from the difference between the model and the observed spectrum
in Fig.~2: both the line strengths and the highly reddened continuum
are well reproduced, except for the continuum level around 5000 \AA
$\,$ where anyway the discrepancy is only at the 4 \% level.
This problem is common to all the models, regardless of the
chosen set of parameters (Figs.~2, 3 and 4).

\hbox{~}
\centerline{\psfig{file=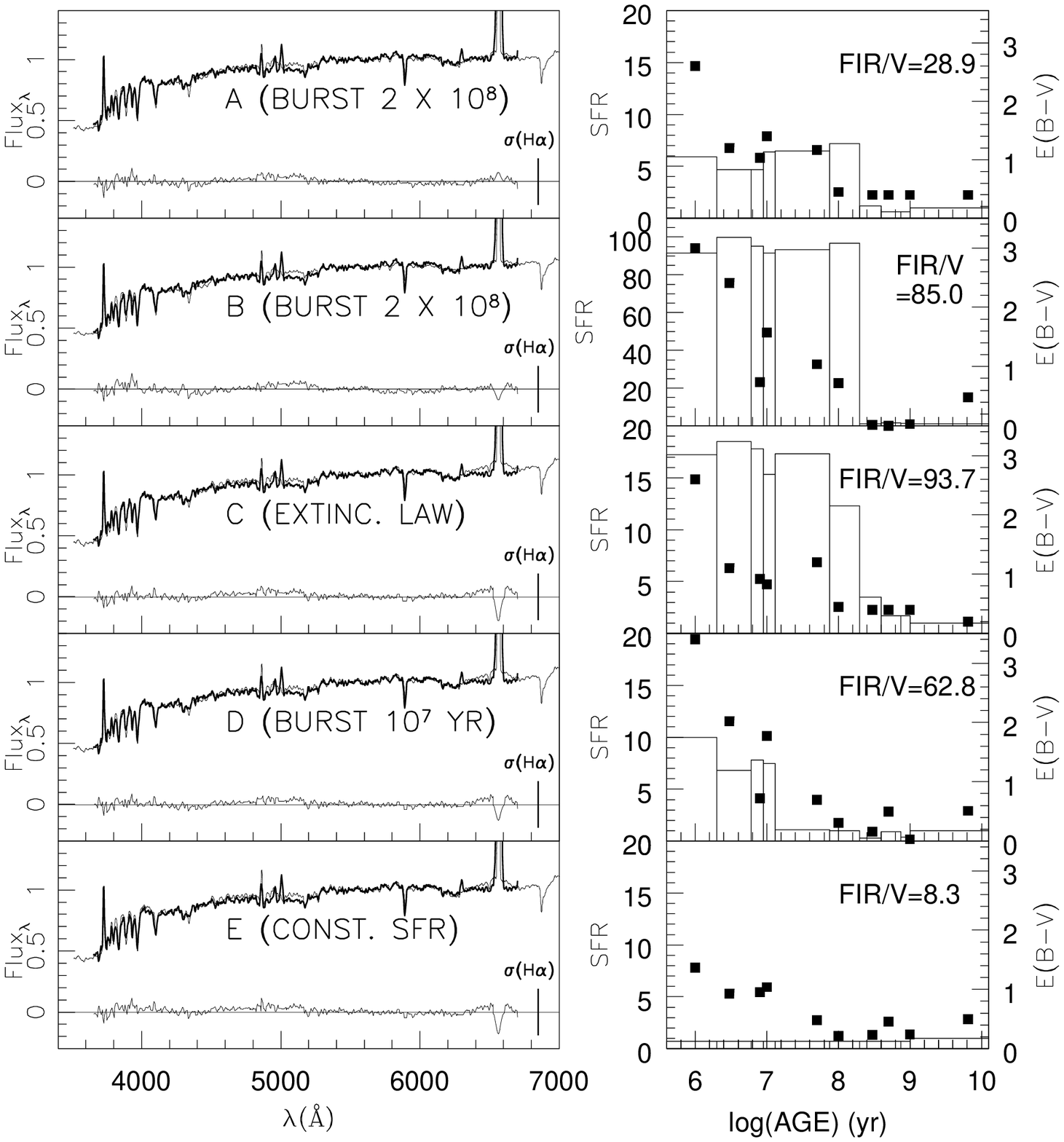,angle=0,width=5in}}
\vspace{-0.4cm}
\noindent{\scriptsize
\addtolength{\baselineskip}{-3pt} 
\hspace*{0.3cm} {\bf Fig.~2.}  {\em Left}: Comparison between the average
observed spectrum of the e(a) galaxies (thick line) and the spectrum
of the starburst 
models listed in Table~1 (thin line) normalized at 5500 \AA. 
The emission lines of the models 
have been assigned a Gaussian shape of (arbitrarily) small width,
hence their equivalent width, not their shape, is relevant for the fit 
(see Table~3).
The difference between the model and the observed spectrum
is also shown. When plotting this difference we have not considered
the emission lines which we did not attempt
to model ([O{\sc III}]4959 and 5007 and [O{\sc I}]6300)
and the NaI D line at 5893 \AA $\,$ which is mostly of interstellar origin.
The vertical segment on the right side represents
the 1 $\sigma$ error in $\rm H\alpha$.
The difference in the flux of the lines has been plotted as a Gaussian
with a width equal to that of the observed line.
{\em Right}: SFR (histogram, normalized to =1 in the old
population) and E(B-V) (dots) of the models whose spectrum is shown in the
left panel. The observed FIR/V ratio is =88.0 (Table~3). 
\addtolength{\baselineskip}{3pt}
}

\vspace{0.2in}
Model A can only account for about 1/3 of the observed
average FIR luminosity ($L_{FIR}/L_V=28.9$ instead of 88.0).
There can be several reasons for this disagreement:
first of all, it should be kept in mind that 
the observed spectra refer only to the central region of the
galaxy as it appears in the optical\footnote{In a typical FIR galaxy,
about 2/3 of the {\em observed} $\rm H\alpha$ luminosity come from regions
more than 1 kpc from the nucleus (Armus et al. 1990). Neverthless, considering
that the color excess measured from the Balmer decrement
strongly decreases with radius (Veilleux et al. 1995, Kim et al. 1998)
and that the primary mechanism for ionizing the emission-line gas outside
of the nucleus could be shock-excitation from outflowing winds
instead of heating by young stars (Armus et al. 1989),
the majority of the {\em intrinsic} $\rm H\alpha$ luminosity (after correcting
for extinction) related to star formation should arise from the 
central regions.}. 
However, also the IR emissivity
is usually concentrated in the central regions of luminous infrared
galaxies (see e.g. Soifer et al. 2000, Kennicutt 1998 and references
therein), therefore aperture effects
are not likely to be responsible for the remaining 2/3 of FIR emission.

Moreover, there might be star forming regions that are
{\it completely} obscured at optical wavelengths.
This is known to be a common situation in Luminous Infrared galaxies
and a spectacular example is the 
image of the Antennae galaxies (Mirabel et al. 1998). The most intense 
starburst in this merging system takes place in a region that is 
inconspicuous at optical wavelengths and
coincides neither with any optically bright region nor with the dark lanes 
observed in the optical image. Such a starburst would give no contribution
to our modelled spectrum, while accounting for a significant fraction
of the FIR luminosity; therefore, our estimate of the current star formation 
rate relative to the old stellar generations would be correspondingly
underestimated.

Model B is an example that describes a starburst with a high FIR/V ratio
due to some practically obscured regions.
As in model A, the starburst began $2 \cdot 10^8$ yr ago, but both the
SFR during the burst and the extinction of the two youngest stellar generations
are higher than in A. The fit of the observed spectrum is 
again remarkable, with a small difference in $\rm H\alpha$ being
well within the 1 $\sigma$ error (Fig.~2). 
Thus, model B can simultaneously account
for the optical properties and the FIR luminosity.

For a standard Salpeter IMF in the range $0.1-100 \, M_{\odot}$, in model
B about 60\% of the total mass in stars is formed during the burst.
Assuming a top-heavy IMF during this phase would substantially reduce
the mass fraction in young stars: for example, with a Salpeter IMF lower mass
limit $=1 \, M_{\odot}$, the starburst in model B forms about 40\%
of the total stellar mass. 
\footnote{The spectral energy distribution of a young (e.g.
$\le 2 \times 10^8$ yr) stellar
generation with an IMF between 0.1 and 100 $M_{\odot}$
is identical to that of a population
with an IMF truncated at low masses (e.g. $m_{inf}=
1M_{\odot}$) because the contribution of stars
with $M<1 M_{\odot}$ is negligible at these ages.
Hence, a model with a standard IMF (any of those presented in this paper)
gives the same results of a model with a truncated IMF during the burst 
with the same set of E(B-V) values
and the same star formation rate {\em in stars more massive than
1 $M_{\odot}$}: the only difference between these two models will be
the {\em total} star formation rate during the starburst, with
the truncated model missing stars below 1 $M_{\odot}$.}

Within the starburst scenario with selective extinction, 
we find an excellent agreement between the model and the observed spectrum
as far as both the continuum shape and the strength of the lines are 
concerned. A model that fits the optical spectrum doesn't necessarily
reproduce the observed FIR/optical ratio but the modelled and observed 
ratios can be reconciled taking into account 
young stellar populations mostly obscured by dust at optical wavelengths.

\subsubsection{Non-standard assumptions regarding the dust: extinction law 
and relative distribution dust-stars}

The models described so far have been computed assuming a standard Galactic
extinction law and a dust screen around each stellar generation.
It is interesting to investigate whether a satisfactory fit can be
found under other assumptions regarding the dust properties.

In model C we consider a $2 \cdot 10^8$ starburst and
an extinction law with $R_V=A_V/E(B-V)=5$ 
as observed towards some dense clouds
in our Galaxy (Mathis 1990). The fit to the e(a) spectrum is still acceptable
but the fit of the continuum around the $\rm H\alpha$ line worsens
and there is a small difference in the $\rm H\alpha$ emission
at the $\sim 1 \sigma$ level (Fig.~2).
In this case the observed FIR/optical ratio is fully reproduced
even with E(B-V) values generally lower than in B
because, for a given E(B-V), the optical flux is much more extinguished.

Finally, we modified the assumptions regarding the relative
distribution of dust and stars: instead of adopting a dust screen,
we examined the case of internal dust, in which dust grains and stellar 
populations are uniformly mixed, for a standard extinction law ($R_V=3.1$). 
In this case, the relation between the observed intensity $I^{}_{\lambda}$
and the intrinsic intensity $I^0_{\lambda}$ is the following:
$I^{}_{\lambda}= I^0_{\lambda} \frac{1-e^{-{\tau}_{sc}(\lambda)}}{{\tau}_{sc}(\lambda)}$ where $\tau_{sc}(\lambda)$ is computed according to Calzetti 
et al. (1994). Interestingly, it was impossible
to find a good fit to the observed spectrum. This is due to the fact
that in the internal dust models 
increasing the obscuration (i.e. the intrinsic optical depth) 
does not yield a corresponding
increase in the {\em reddening} of the spectrum: the latter saturates
to an asymptote, and the asymptotic reddening value (E(B-V)$\sim 0.18)$) 
is too low to be able to account for the observed emission line
ratios (E(B-V)$\sim 1.1$ from the observed Balmer decrement, W98a). 
This saturation of the mixed case was noted by Calzetti et al. 
(1994, Sec.5.3) and our asymptotic value of E(B-V)
is in agreement with their asymptotic value of the measured
Balmer optical depth (${\tau}^l_B=0.24$). 
This result can be easily understood considering that 
the most heavily reddened stars
are those most heavily obscured, hence
they cannot provide the dominant contribution to
the emerging light, which instead must come 
from the outermost layers (Witt et al. 1992, Calzetti 1998): 
in the internal dust model
this effect gives rise to a natural limit to the reddening.
While the existence of regions where dust and stars are mixed
can remove the contribution to the optical spectrum of entire stellar 
populations (those most deeply embedded), these regions do
not appear to determine the characteristics
of the {\em emerging} spectrum, which requires a higher
reddening at least partly due to foreground dust. 

\subsubsection{Burst duration and strength}

\noindent
{\it How long (short) can the burst be?}
It is interesting to investigate whether 
an e(a) spectrum (in particular, its
$\rm H\delta$ strength) requires a quite long starburst involving 
intermediate-age populations ($10^8-10^9$ yr).
Contrary to the most intuitive expectations, this is not the case.
As an example of this, we present the good fit obtained for
a starburst that began only $\sim 10^7$ yr ago (model D). 
In this case the strong $\rm H\delta$ line is not produced by the stars
born in the starburst event, but by the previous stellar generations.
We find that
starbursts as short as a few times $10^6$ yr and as long as $10^9$ yr
are able to produce the e(a) spectrum as long as:
a) the youngest populations
($\leq$ a few $10^7$ yr) must be highly extincted and b)
there must be a contribution of the 
``old'' population, but this should not overwhelm the
intermediate-age contribution at $\sim 4000$ \AA.

Model D has a FIR/V=62.8 therefore accounts for more than
2/3 of the observed FIR luminosity, but
as for longer starbursts (models A and B), a good fit to the optical
spectrum can be obtained with short starburst
models spanning a wide range in FIR/V
ratio, simply modifying the amount of ``hidden star formation''.

As an extreme example of what we have just shown and to further
illustrate the degeneracy in the physical solutions allowed by optical
data, we consider a galaxy
with a practically constant SFR during its whole evolution (model E)
and we show that this can still fit the e(a) spectrum reasonably well 
-- with a slight mismatch in $\rm H\alpha$ and in the continuum around it 
(Fig.~2) --
if we adopt high extinction values for the young populations and 
a quite strong extinction of the old population (E(B-V)=0.5).
It is worth stressing that the E model does not represent 
a ``spiral-like'' star formation history, because in all but the latest
spiral types the SFR at recent times is known to be
lower than the mean SFR during the previous evolution (Kennicutt 1998).
We also tried to fit the e(a) spectrum assuming a spiral-like
history in which the SFR of the 
old population is at least 1.5 times the SFR of the young and 
intermediate populations. This model cannot fit the observed
spectrum unless we allow the extinction
of the ``old'' population to be unphysically high, higher than that of the
intermediate-age populations and much higher than the average
observed even in the HII regions of nearby spiral galaxies
(about 1 mag at $\rm H\alpha$, corresponding to E(B-V)$\sim 0.4$).

Although model E is able to reproduce the optical spectrum, its FIR/V is
a factor of 10 lower than what is observed, and in this case (i.e. for an
approximately constant star formation) no
young obscured population can be invoked to explain the missing 
FIR flux: any additional current star formation would qualify such a 
model as a starburst.

\hbox{~}
\centerline{\psfig{file=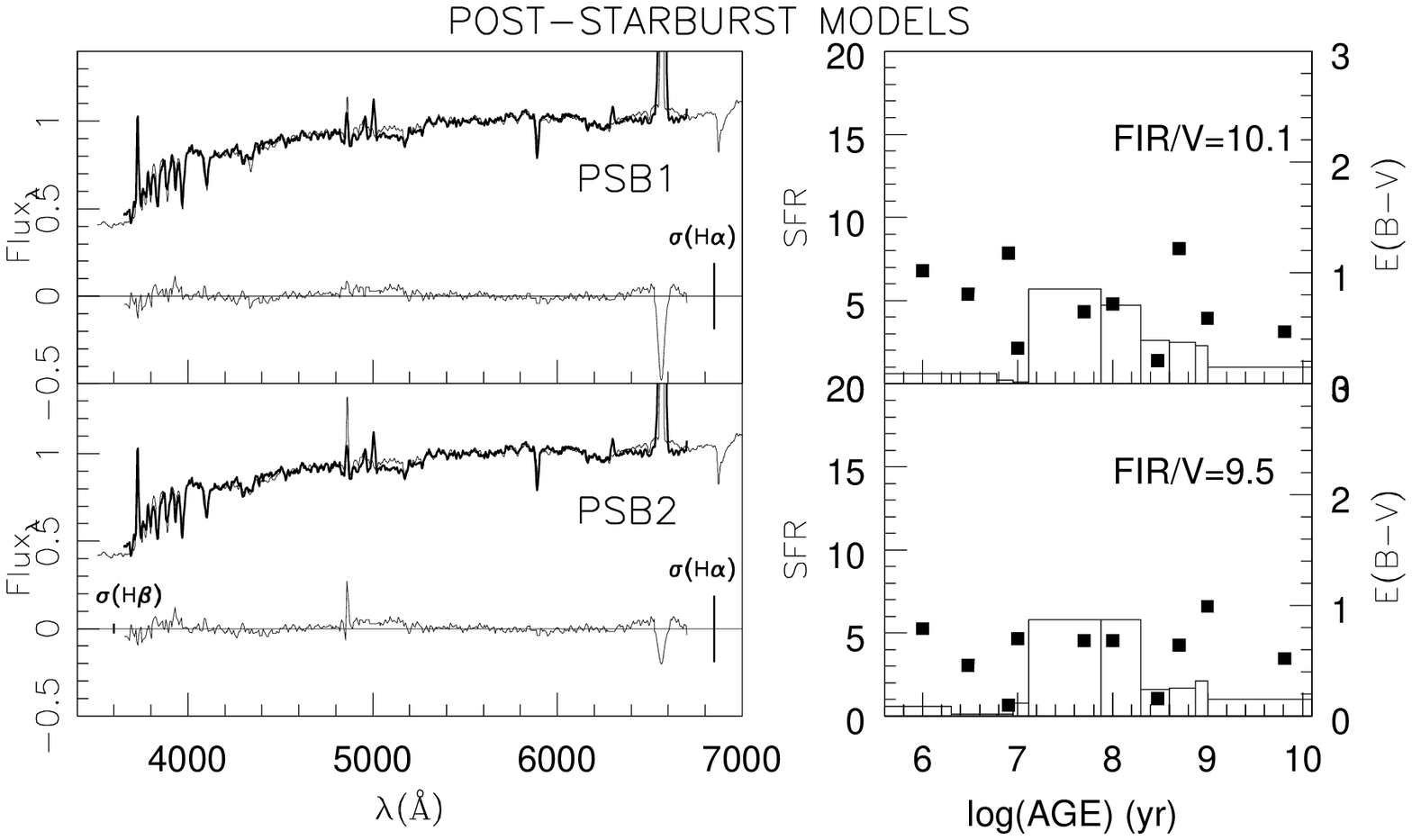,angle=0,width=5in}}
\vspace{-2.0in}
\noindent{\scriptsize
\addtolength{\baselineskip}{-3pt} 
\hspace*{0.3cm} {\bf Fig.~3.} Results of the post-starburst 
models listed in Table~2. Symbols as in Fig.~2. In the lower panel,
the 1 $\sigma$ errorbar in $\rm H\beta$ is shown on the left side.
\addtolength{\baselineskip}{3pt}
}

\subsection{The post-starburst hypothesis}
Here we consider another family of models, in which the galaxy
is seen after a strong starburst during the post-starburst phase
(Table~2 and Fig.~3).
As discussed in P99, the two spectral features that 
define an e(a) spectrum (a strong $\rm H\delta$
line and a moderate [O{\sc ii}]) can be reproduced 
by dust-free models of a post-starburst galaxy with some residual star 
formation, but on the basis of the FIR properties of the e(a) galaxies
and of their unusually low [O{\sc ii}]/$\rm H\alpha$ ratios,
the same authors concluded that e(a) spectra more likely belong to starburst
galaxies that are highly extincted by dust.\footnote{ 
The post-starburst interpretation has been mentioned by Flores et al. (1999)
(see \S2),
who consider their data to be consistent with the last phases of a burst
following strong star formation episodes, where the MIR emission is still high
due to dust heating by intermediate-mass stars ($M=1-3 \, M_{\odot}$).}
Here we contemplate post-starburst models with a small residual star
formation activity and including extinction by dust.
This family of models shows a tendency either to underestimate
the $\rm H\alpha$ strength or to overestimate $\rm H\beta$ (see models
PSB1 and PSB2 in Fig.~3 and in Table~3): the emission in 
at least one of these two lines
is more than 1 $\sigma$ away from the observed value.
The reason for this is that the observed $\rm H\alpha/H\beta$ ratio
requires a high extinction of the lines (the extinction 
measured from this line ratio is E(B-V)$\sim 1.1$),
but such a high extinction in a post-starburst galaxy
would produce too weak emission lines i.e. there isn't enough current 
star formation to give the observed strength.
Although the discrepancy between
the model and the observed $\rm H\beta$ strength in model PSB2
is not dramatic on an absolute scale (still less than 3 \AA),
the post-starburst interpretation is not favoured even on the basis of 
the optical data alone (without taking into account the strong FIR emission)
{\it if the whole spectrum including the $\rm H\alpha$ line} is examined.
In addition, the FIR luminosity produced by these models
is just a small fraction of the observed value and, similarly to model E,
within the post-starburst scenario
no young obscured population can be invoked to explain this
discrepancy.

\begin{table*}
{\scriptsize
\begin{center}
\centerline{\sc Table 1: Populations of the starburst models}
\vspace{0.1cm}
\begin{tabular}{lcccccccccc}
\hline\hline
\noalign{\smallskip}
Age & \multispan{2}{\hfil{Model A}\hfil} & 
\multispan{2}{\hfil{Model B}\hfil} & 
\multispan{2}{\hfil{Model C}\hfil} & \multispan{2}{\hfil{Model D}\hfil} & 
\multispan{2}{\hfil{Model E}\hfil} \cr
yr & 
\multispan{2}{\hfil{$R_V=3.1$}\hfil} & 
\multispan{2}{\hfil{$R_V=3.1$}\hfil} & 
\multispan{2}{\hfil{$R_V=5$}\hfil} & 
\multispan{2}{\hfil{$R_V=3.1$}\hfil} &
\multispan{2}{\hfil{$R_V=3.1$}\hfil}  \cr
\noalign{\smallskip}
 & SF & E(B-V) & SF & E(B-V) & SF & E(B-V) & SF & E(B-V) & SF & E(B-V) \cr
\hline
\noalign{\medskip}
$10^6$          &  5.9&  2.60&  91.5& 3.00& 17.2&  2.60& 10.0&  3.40&  0.7&  1.37 \cr
$3 \cdot 10^6$  &  4.7&  1.20& 100.0& 2.41& 18.5&  1.10&  6.8&  2.02&  0.7&  0.93 \cr 
$8 \cdot 10^6$  &  4.7&  1.03&  95.3& 0.74& 17.8&  0.92&  7.8&  0.72&  0.7&  0.95 \cr
$10^7$          &  6.4&  1.40&  91.5& 1.57& 15.3&  0.83&  7.5&  1.77&  0.7&  1.04 \cr
$5 \cdot 10^7$  &  6.5&  1.17&  93.4& 1.04& 17.3&  1.20&  1.1&  0.70&  0.7&  0.48 \cr
$10^8$          &  7.2&  0.45&  96.8& 0.72& 12.3&  0.45&  1.0&  0.31&  0.7&  0.22 \cr
$3 \cdot 10^8$  &  1.2&  0.40&  1.0&  0.02&  3.5&  0.40&  0.3&  0.16&  0.7&  0.23 \cr
$5 \cdot 10^8$  &  0.6&  0.40&  1.5&  0.00&  1.7&  0.40&  0.9&  0.50&  0.7&  0.46 \cr
$10^9$          &  0.6&  0.40&  1.3&  0.03&  1.7&  0.40&  0.4&  0.03&  0.7&  0.24 \cr
old             &  1.0&  0.40&  1.0&  0.48&  1.0&  0.20&  1.0&  0.51&  1.0&  0.50 \cr
\noalign{\smallskip}
\noalign{\hrule}
\noalign{\smallskip}
\end{tabular}
\end{center}
}
\vspace*{-0.8cm}
\end{table*}

\begin{table*}
{\scriptsize
\begin{center}
\centerline{\sc Table 2: Populations of the post-starburst (PSB)
and constant extinction (CE) models}
\vspace{0.1cm}
\begin{tabular}{lcccccccc}
\hline\hline
\noalign{\smallskip}
Age & 
\multispan{2}{\hfil{Model PSB1}\hfil} 
& \multispan{2}{\hfil{Model PSB2}\hfil} &
\multispan{2}{\hfil{Model CE1}\hfil} 
& \multispan{2}{\hfil{Model CE2}\hfil} \cr
yr  & 
\multispan{2}{\hfil{$R_V=3.1$}\hfil} & \multispan{2}{\hfil{$R_V=3.1$}\hfil} &
\multispan{2}{\hfil{$R_V=3.1$}\hfil} & \multispan{2}{\hfil{$R_V=3.1$}\hfil} 
\cr
\noalign{\smallskip}
 & 
SF & E(B-V) & SF & E(B-V) &
SF & E(B-V) & SF & E(B-V) 
\cr
\hline
\noalign{\medskip}
$10^6$          &  0.6&  1.02&  0.6&  0.79&  0.05& 0.56&  8.3 & 0.56 \cr
$3 \cdot 10^6$  &  0.6&  0.81&  0.1&  0.46&  0.2&  0.56&  8.3 & 0.56 \cr 
$8 \cdot 10^6$  &  0.2&  1.18&  0.1&  0.10&  4.5&  0.56&  8.6 & 0.56 \cr
$10^7$          &  0.1&  0.32&  0.8&  0.70&  4.4&  0.56&  8.3 & 0.56 \cr
$5 \cdot 10^7$  &  5.7&  0.65&  5.8&  0.68&  13.6& 0.56&  13.1& 0.56 \cr
$10^8$          &  4.7&  0.72&  5.8&  0.68&  1.4&  0.56&  0.1 & 0.56 \cr
$3 \cdot 10^8$  &  2.6&  0.21&  1.6&  0.16&  11.5& 0.56&  22.2& 0.56 \cr
$5 \cdot 10^8$  &  2.5&  1.22&  1.7&  0.64&  3.8&  0.56&  22.3& 0.56 \cr
$10^9$          &  2.3&  0.59&  2.1&  0.99&  0.1&  0.56&  18.9& 0.56 \cr
old             &  1.0&  0.47&  1.0&  0.52&  1.0&  0.56&  1.0 & 0.56 \cr
\noalign{\smallskip}
\noalign{\hrule}
\noalign{\smallskip}
\end{tabular}
\end{center}
}
\vspace*{-0.8cm}
\end{table*}

\begin{table*}
{\scriptsize
\begin{center}
\centerline{\sc Table 3: Results}
\vspace{0.1cm}
\begin{tabular}{lccccccl}
\hline\hline
\noalign{\smallskip}
   & W($\rm H\alpha$+NII)
$^a$ & W($\rm H\beta$) & W($\rm H\delta$) & W([O{\sc ii}]) & 
$L_{FIR}/L_V$$^b$ & SFR$^c$ & description \cr
\hline
\noalign{\medskip}
 Wu$^d$  & -62.48$\pm$5.3 & 0.68$\pm$0.8 & 5.64$\pm$0.5 & -12.68$\pm$2.1 & 
88.0$\pm$13.0 & -- & observed e(a) spectrum \cr
\noalign{\smallskip}
\noalign{\hrule}
\noalign{\smallskip}
\multispan{8}{\hfil{STARBURST~MODELS}\hfil}\cr
\noalign{\smallskip}
Model A & -65.05 & 0.72 & 5.52 & -11.40 & 28.9 & 38 & $2 \cdot 10^8$ yr burst \cr
Model B & -61.26 & 0.30 & 5.65 & -11.69 & 85.0 & 102 & $2 \cdot 10^8$ yr burst, high FIR/5500 \cr
Model C & -56.97 & 0.72 & 5.13 & -12.71 & 93.7 & 117 & different extinction law \cr
Model D & -59.46 & 0.43 & 5.33 & -12.23 & 62.8 & 126 &  $10^7$ yr burst \cr
Model E & -59.03 & 0.55 & 5.72 & -12.24 & 8.3  & 9 & $\sim$ constant SFR \cr
\noalign{\smallskip}
\noalign{\hrule}
\noalign{\smallskip}
\multispan{8}{\hfil{POST-STARBURST~MODELS}\hfil}\cr
\noalign{\smallskip}
Model PSB1 & -46.50 & 0.71 & 5.81 & -12.36 & 10.1 &  4 & post-SB 1 \cr
Model PSB2 & -56.68 & -2.06 & 5.25 & -13.05 & 9.5 & 4 & post-SB 2 \cr
\noalign{\smallskip}
\noalign{\hrule}
\noalign{\smallskip}
\multispan{8}{\hfil{MODELS~WITH~CONSTANT~EXTINCTION}\hfil}\cr
\noalign{\smallskip}
Model CE1  & -37.64 & -0.44 & 4.94 & -15.49 & 9.7 & 0.2 & Constant ext., free SF \cr
Model CE2  & -379.54 & -64.47 & -6.97 & -152.86 & 12.3 & 13 & Constant ext., SB model \cr
\noalign{\smallskip}
\noalign{\hrule}
\noalign{\smallskip}
\multispan8{a) All equivalent widths are in \AA, negative in emission and 
positive in absorption.
The model W($\rm H\alpha$+NII) was found \hfil}\cr 
\multispan8{adopting 
$NII=0.2 \rm H\alpha$ which is the mean value observed by Tresse
et al. (1999) for this $\rm H\alpha$ strength. \hfil}\cr
\multispan8{b) $L_{FIR}/L_V$=$L_{FIR}/(5500 \times L_{5500})$
where $L_{FIR}$ ($\rm erg \, s^{-1}$) is the total FIR luminosity
and $L_{5500}$ ($\rm erg \, s^{-1} \, {\AA}^{-1}$) \hfil}\cr 
\multispan8{is the luminosity
at 5500 \AA $\,$ as derived from the spectrum.\hfil}\cr 
\multispan8{c) Current star formation rate in $M_{\odot} \, yr^{-1}$
derived from the model fitting the observed luminosity at 5500 \AA
$\,$ assuming \hfil}\cr
\multispan8{a Salpeter IMF (x=2.35, 0.1-100 $M_{\odot}$).
\hfil}\cr
\multispan8{d) Errors quoted are standard deviations of the bootstrap median
value within the sample of 19 galaxies. \hfil}\cr
\noalign{\smallskip}
\end{tabular}
\end{center}
}
\vspace*{-0.8cm}
\end{table*}

\subsection{A reference case: models with constant extinction}
So far we have only considered models in which the E(B-V) is allowed to 
vary from one stellar population to another. In this section we
contemplate more ``canonical'' models which have a common value of
extinction for all the stellar generations, showing that these fail
to reproduce the e(a) spectrum.

Having imposed the extinction to be the same for all the 10 populations
and to be within the range $0<E(B-V)<3$ mag, the model was let 
free to find a fit without imposing any constraint on the star formation
rate of the various generations. The ``best fit'' model obtained in this
way (CE1) is shown
in the top panel of Fig.~4 and in Tables~2 and 3: while the shape of 
the continuum can be fitted with an E(B-V)=0.56, 
this extinction cannot reproduce the equivalent widths
of the emission lines and the model $\rm H\alpha$ line is too weak compared
to the one observed.

For comparison, we also present a starburst model with a constant
extinction (model CE2, lower panel of Fig.~4, Tables~2 and ~3): having imposed
a current burst\footnote{This was done imposing that the SFR during
the last $10^7$ was at least twice as much the maximum
SFR allowed in the old population.}, ``the best fit''
model can reproduce the continuum with the same extinction as model CE1, 
but presents far too strong emission lines.
In conclusion, while a fit to the continuum can be found with an E(B-V)
at most $\sim 0.6$ (as shown by these models with constant
extinction), the line ratios necessitate a color excess {\em greater}
than 1, as proved by the observed Balmer decrement.
Selective extinction is required to reproduce simultaneously
the strength of the lines and the shape of the continuum of e(a)
spectra.

\hbox{~}
\centerline{\psfig{file=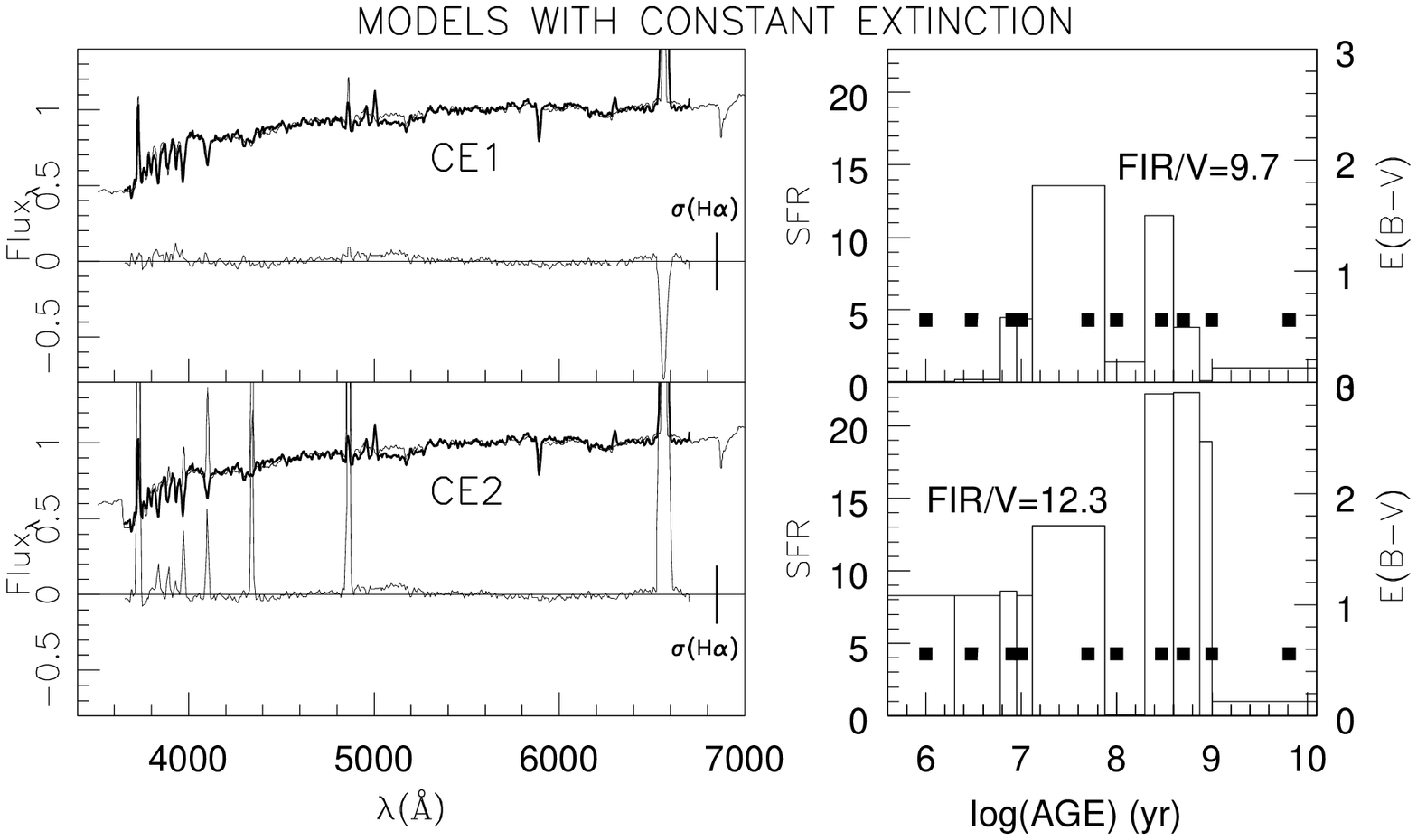,angle=0,width=5in}}
\vspace{-2.0in}
\noindent{\scriptsize
\addtolength{\baselineskip}{-3pt} 
\hspace*{0.3cm} {\bf Fig.~4.} Results of the models with constant
extinction listed in Table~2. Symbols as in Fig.~2.
\addtolength{\baselineskip}{3pt}
}

\section{Conclusions}

Spectra with a strong $\rm H\delta$ line in absorption and a moderate
[O{\sc ii}] line in emission -- named e(a) spectra -- are numerous
among luminous infrared galaxies. We have shown that the optical 
e(a) spectrum can be well reproduced by
a spectrophotometric model that includes both the stellar component
and the emission lines and continuum of the ionized gas,
{\it assuming dust extinction to vary with the age of the stellar
population}. 

The main conclusions of this work can be summarized as follows:

1) On the basis of our models, 
two conditions are necessary to explain the e(a) spectra:

a) the youngest stellar generations (age $\leq 10^7$) in e(a) galaxies
must be highly obscured by dust. Their extinction values are
significantly higher than those applying to the previous stellar populations.
Models with a constant value of extinction are unsuccessful in  
fitting the optical spectrum;

b) these galaxies must host strong ongoing star formation at a level 
higher than in quiescent spirals. Post-starburst models with a small residual
star formation fail to explain simultaneously the strength
of the $\rm H\alpha$ and of the $\rm H\beta$ lines. 

2) The degeneracy of the optical spectrum hinders a quantitative estimate 
of the burst duration and intensity. 
The spectrum is consistent with, but does not require, a ``long'' starburst 
that began $10^8-10^9$ yr ago. 

3) A model fitting the e(a) spectrum does not necessarily produce
the FIR/optical ratio observed: the optical spectrum (including
the observed $\rm H\alpha$ luminosity and $\rm H\alpha/H\beta$ ratio)
does not constrain the current star formation rate.
The observed FIR/V can be explained
by starburst models if a significant fraction of the FIR luminosity
originates in regions that are practically 
obscured at optical wavelengths.

4) All the results listed above refer to the case of a dust screen
placed in front of each stellar generation.
Models in which dust and stars are uniformly mixed fail to fit
the e(a) spectrum, producing too low a reddening of the emerging
emission lines. The observed reddening seems to be
at least partly due to foreground dust. 

5) The extinction correction factor at $\rm H\alpha$ 
found from the Balmer decrement in the e(a) Very Luminous Infrared galaxies
of our sample ($E(\rm H\alpha)\sim 15$, 
corresponding to a median E(B-V)=1.1, W98a) is in agreement
with that derived with the same method in other FIR luminous samples
(Kim et a. 1998, Veilleux et al. 1995, median E(B-V) for non-AGN
galaxies =1.13 and 1.05, respectively). This factor is about 6 times
higher than the average $E(\rm H\alpha)=2.5$ in nearby spirals
(Kennicutt 1992).
Ignoring the fact that the slit might have missed a fraction of the 
$\rm H\alpha$ luminosity and applying the extinction 
correction to derive the total current star formation rate from the relation
$\rm SFR(M_{\odot} \, yr^{-1}) = 0.8 \times 10^{-41} \, E(\rm H\alpha)) 
\times L(H\alpha) \, erg \, s^{-1} $ (Kennicutt 1992)\footnote{For 
a Salpeter IMF (x=2.35, $0.1-100 M_{\odot}$) and a standard Galactic 
extinction law ($R_V=3.1$).}, yields a value of the SFR that
is still underestimated
by a factor $\sim 3$ compared to the SFR derived from the FIR luminosity.

\section*{Acknowledgements} 
We are grateful to Hong Wu for providing us
the spectral catalog in a convenient format,
and to Lee Armus, Herv\'e Aussel, Rob Kennicutt, Pasquale Panuzzo,
Ian Smail and Hong Wu for their help
and/or comments regarding the manuscript.
We sincerely thank the anonymous referee for 
constructive suggestions that improved this paper.
BMP and AF thank the organizers of the 2000 Ringberg
Starburst meeting, where some of the issues in this
paper were discussed.
BMP acknowledges a discussion (Cambridge (UK), 1997) 
with Anne L. Kinney, who pointed out the need for a
spectrophotometric model with selective extinction.

\smallskip

\end{document}